\documentclass[aps,prl,twocolumn,showpacs,superscriptaddress,groupedaddress]{revtex4}  
\usepackage{dcolumn}   
\usepackage{bm}        
\usepackage{amssymb}   
\usepackage{amsmath}

\usepackage[english]{babel}
\usepackage{pgfplots}
\usepackage{textcomp}
\usepackage{color}

\usepackage[T1]{fontenc}
\usepackage{siunitx}
\usepackage{graphicx}
\usepackage{dcolumn}
\usepackage{bm}

\begin{document}


\title{Near-surface rheology and hydrodynamic boundary condition \\
of semi-dilute polymer solutions}


\author{Gabriel~Guyard}
\affiliation{Gulliver CNRS UMR 7083, PSL Research University, ESPCI Paris, 10 rue Vauquelin, 75005 Paris, France}
\affiliation{Universit\'e Paris-Saclay, CNRS, Laboratoire de Physique des Solides, 91405, Orsay, France}

\author{Alexandre~Vilquin}
\affiliation{Gulliver CNRS UMR 7083, PSL Research University, ESPCI Paris, 10 rue Vauquelin, 75005 Paris, France}

\author{Nicolas~Sanson}
\affiliation{Laboratoire SIMM, CNRS UMR 7615, PSL Research University, ESPCI Paris, 10 rue Vauquelin, 75005 Paris, France}
\affiliation{Laboratoire Physico-chimie des Interfaces Complexes, ESPCI Paris, 10 rue Vauquelin, F-75231 Paris, France}

\author{Fr\'ed\'eric~Restagno}
\affiliation{Universit\'e Paris-Saclay, CNRS, Laboratoire de Physique des Solides, 91405, Orsay, France}

\author{Joshua~D.~McGraw}
\affiliation{Gulliver CNRS UMR 7083, PSL Research University, ESPCI Paris, 10 rue Vauquelin, 75005 Paris, France}





\date{\today}

\begin{abstract}
Understanding confined flows of complex fluids requires simultaneous access to the mechanical behaviour of the liquid and the boundary condition at the interfaces. Here, we use evanescent wave microscopy to investigate near-surface flows of semi-dilute, unentangled polyacrylamide solutions. By using both neutral and anionic polymers, we show that monomer charge plays a key role in confined polymer dynamics. For solutions in contact with glass, the neutral polymers display chain-sized adsorbed layers, while a shear-rate-dependent apparent slip length is observed for anionic polymer solutions. The slip lengths measured at all concentrations collapse onto a master curve when scaled using a simple two-layer depletion model with non-Newtonian viscosity. A transition from an apparent slip boundary condition to a chain-sized adsorption layer is moreover highlighted by screening the charge with additional salt in the anionic polymer solutions. We anticipate that our study will be a starting point for more complex studies relating the polymer dynamics at interfaces to their chemical and physical composition.
\end{abstract}

\maketitle

\section{Introduction}
\begin{figure*}[t]
\centering
\includegraphics[scale=1]{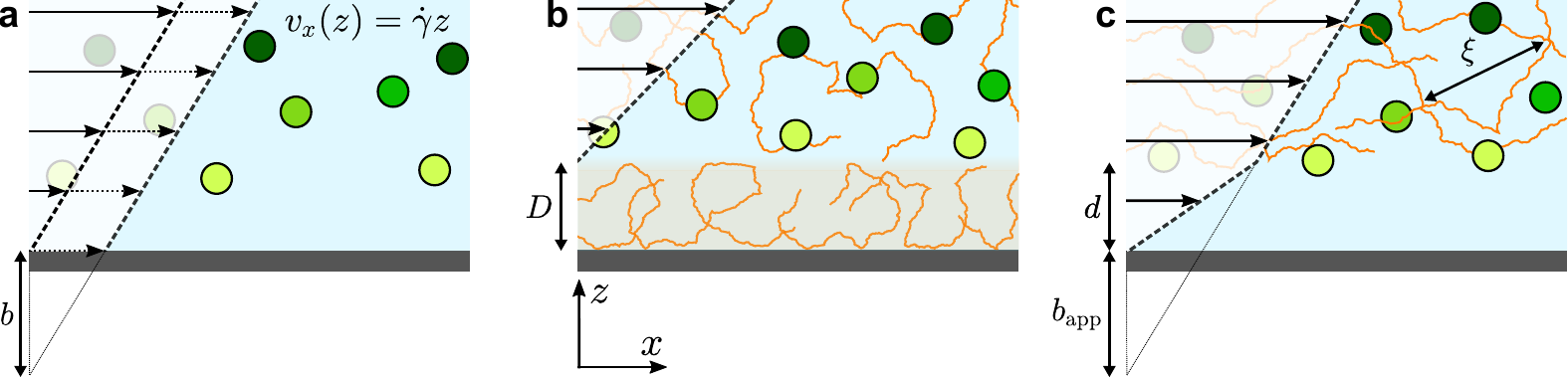} 
\caption{Schematics of different near-wall velocity profiles and associated boundary conditions. (\textbf{a}) The classic no-slip velocity profile (solid arrows) typically observed for simple liquids at large scales; also shown is  a velocity profile with a non-zero slip velocity (dashed arrows), and the slip length, $b$. (\textbf{b}) Adsorbed chains with hydrodynamically arrested regions of macromolecular size, $D$, displacing the no-slip plane upward. (\textbf{c}) The depletion layer case, with apparent slip due to a viscosity mismatch between the two sub-layers. }
\label{introduction_schem}
\end{figure*}

The physics of polymers near interfaces is encountered in many fundamental and applied problems, from industrial to biological processes. In this context, interfacial interactions are a key feature.~\cite{de_gennes_polymer_1981} Attractive surfaces, for example, may lead to irreversible adsorption of polymer chains,~\cite{FU1998, leger_surface-anchored_1999} but chains may also be repelled from surfaces~\cite{lee1991direct} leading to a depletion layer. Such structural effects can have major consequences on dynamics in confined systems. Adsorption leads to decreased permeability in porous media, which is a major issue in enhanced oil recovery;\cite{bessaies-bey_impact_2018} depletion enhances flows via the formation of a low-viscosity lubrication layer close to the surface.~\cite{Barnes1995, Muller1990} As will be studied here, polymer/surface interactions impact nanoscale flows by modifying the boundary condition for polymeric fluids, and non-trivial rheology also plays a key role. 

The classic, no-slip boundary condition, schematically depicted in Fig.~\ref{introduction_schem}(a), assumes that the hydrodynamic velocity profile vanishes at the wall. While this condition is the simplest to treat analytically, there is no \emph{a priori} reason for its validity. Slip at the wall was thus hypothesised in the early stages of hydrodynamics,~\cite{navier1823memoire} and the velocity at the wall is typically characterised by a slip length, $b$, defined as the distance at which the flow profile linearly extrapolates to zero (see Fig.~\ref{introduction_schem}(a)). For simple liquids on hydrophobic surfaces, slip lengths are just a few nanometres.~\cite{neto_boundary_2005, Lauga2007, bocquet_nanofluidics_2010} For complex fluids,~\cite{hatzikiriakos_slip_2015} however, these extrapolation lengths can reach many microns; this is notably the case for polymer melts and solutions.~\cite{mhetar_slip_1998} 

Slip of polymer melts has indeed been extensively studied over several decades,~\cite{leger_wall_1997, Baumchen2012, HATZIKIRIAKOS2012} and the case of ideal surfaces was explained in the seminal paper of de Gennes.~\cite{deGennes1979} There, arguing from the standpoint of stress balance at the interface, a cubic dependence on the molar mass of the slip length was predicted for smooth non-adsorbing surfaces. This prediction was confirmed recently in studies of dewetting thin polymer films~\cite{baumchen_reduced_2009} and using velocimetry measurements.~\cite{henot_friction_2018,Henot2018} For adsorbing surfaces or surfaces with grafted chains, the field is still active \cite{Ilton2018,HamiehSubstrateDewetting2007} but a mature understanding both on the experimental~\cite{leger_wall_1997} and theoretical sides~\cite{Brochard1996} has been reached to capture the main picture.  
The dynamics of dilute polymer solutions near interfaces has also received significant attention.~\citep{graham_fluid_2011} In this situation, the chains can be considered as independent. The typical relevant force is a repulsive hydrodynamic interaction with the wall under flow,~\cite{Chen2005} along with molecular forces which can lead to complex relationships between the shear rate and a depletion layer at the wall.\cite{Park2019} 

In between the extremes of dilute polymer solutions and polymer melts, the semi-dilute solution regime is encountered in which polymer chains overlap, but where the solvent still plays an important role. For these cases, relatively little is known about the boundary conditions and near-wall transport.~\cite{graham_fluid_2011} On one hand, attractive interactions can lead to adsorbed layers as described above, leading to an immobile layer of thickness $D$ (Fig.~\ref{introduction_schem}(b)). Such a layer can be interpreted as providing a ``negative'' slip length. On the other hand, for repulsive surfaces, apparent slip lengths of many micrometers~\cite{Muller1990} are thought to arise due to the depletion layers schematically indicated in Fig.~\ref{introduction_schem}(c). More recent experimental studies~\cite{cuenca_submicron_2013} showed suppressed apparent slip for polymers confined to regions smaller than the bulk chain size, which was recovered lately in surface forces apparatus experiments of entangled, semi-dilute solutions.~\cite{barraud_large_2019} Relatively few studies have been performed on semi-dilute, unentangled polymer solutions where the slip lengths may be smaller. Indeed, some of the most recent velocimetry techniques for the investigation of near-wall polymer dynamics~\cite{Henot2017,Grzelka2020} are limited to micron-scale slip lengths. As will be shown below, however, there is a rich phenomenology below micrometric scales.

Semi-dilute polymer solutions, both entangled and unentangled, are also commonly observed to be non-Newtonian fluids.~\cite{Larson1999} Such fluids particularly display shear viscosities that decrease with the imposed shear rate.~\cite{colby_shear_2007} In the bulk and for neutral polymers in good solvents, universal behaviour for this non-linearity has been demonstrated for a variety of polymer/solvent pairs;~\cite{Heo2008, Jouenne2020} however, it is not obvious that such universality should be preserved near the interface. A consequence of this non-trivial rheological behaviour is that slip measurements should be accompanied with simultaneous rheometry measurements, and that this rheology should be done near the surface. Indeed, when the complex interfacial phenomena occur, usual macroscopic rheometry tools could be subject to subtle effects that make the interpretation of data difficult.~\cite{kiljanski_method_1989, Barnes1995, bertula_strain-stiffening_2019} This sets the need for experimental setups able to distinguish near-surface rheology from the boundary condition. One way to achieve both at the same time is to map the polymer flow near the wall using particle-tracking velocimetry.~\cite{wang_homogeneous_2011, Nghe2011}

In this paper, we use total internal reflection fluorescence microscopy (TIRFM) to perform 3-dimensional velocimetry in semi-dilute unentangled polyacrylamide (PAM) solutions. We recover the near-wall velocity profiles of PAM solutions and capture their shear-thinning behaviour in good agreement with bulk rheometry. In the same experiment, we measure complex boundary conditions that depend on the presence of polymer chains, their charge and the electrolyte concentration. We demonstrate that neutral polymer chains adsorb at the surface of the channel, while solutions containing negatively-charged chains display apparent slip at the glass wall with a shear rate dependent slip length ranging between a few and 2000\ \SI{}{\nano\meter}.
\section{Materials and methods}

\subsection{Polymers, microfluidics, and rheometry}

The polymers used in this study were neutral and anionic, monodisperse PAMs synthesised by controlled-radical polymerisation as described elsewhere.~\cite{Read2014, BESSAIESBEY2019} For anionic polymers, negative charges were introduced by 10\% random copolymerisation of acrylic acid. Table~\ref{polymers} lists the characteristics and nomenclatures for all of the polymers used here. 

Aqueous solutions were prepared by dissolving PAM in ultrapure deionised water (18.2\ \SI{}{\mega\ohm\per\centi\meter}, Milli-Q Advantage A10) at different mass fractions ranging from $c = 2$ to $10$\  \SI{}{\milli\gram\per\milli\liter} 
for neutral PAM and from $0.1$ to $2$\ \SI{}{\milli\gram\per\milli\liter} for the anionic polymers; these concentrations can be compared to our estimates of the overlap concentration,  $c^*$, of the different polymers in water (see Table~\ref{polymers}). The overlap concentrations for the neutral PAMs were evaluated based on the hydrodynamic radius of the neutral polymers ($50$\ \SI{}{nm} and $66$\ \SI{}{nm} for PAM(1284k) and PAM(2082k)) and standard scaling laws for polyelectrolyte chains.~\cite{Dobrynin1995} For the anionic polymers, we also performed experiments with added salt, using $28$\ mM NaCl (Sigma Aldrich). For viscous, Newtonian liquids, deionised water/glycerol mixtures (Sigma Aldrich, $\geq 99\%$) with composition between $0$ and $60$\ wt.\% glycerol were used. Finally, in order to probe the near-surface flow using TIRFM, $110$ nm-diameter carboxylate-modified fluorescent microspheres (Invitrogen F8803, $2$\ \% solid) were added to the PAM solutions as received by the manufacturer at a volume fraction of $3$\ \SI{}{\micro\liter\per\milli\liter}.

\begin{table}[b]
\small
  \caption{\ Characteristics of PAMs used in this study}
  \label{polymers}
  \begin{tabular*}{0.48\textwidth}{@{\extracolsep{\fill}}lllll}
    \hline
    & & & & Electric \\
   Designation &$M_\mathrm{w}$ (\SI{}{\kilo\gram \per \mol}) & $c^*$ (\SI{}{\milli\gram\per\milli\liter}) & \DJ & charge \\
   
    \hline
    
    PAM(1284k) & \num{1\,284} & $4.0$ & $1.05$ & neutral \\
    PAM(2082k) & \num{2\,082} & $2.8$ & $1.07$ & neutral \\
    PAM(817k$^{[-]}$) & \num{817} & $1.5\times10^{-2}$ & $1.09$ & $[-]$, $10$\ \%  \\
    \hline
  \end{tabular*}
\end{table}

The tracer-containing solutions described above were injected in a microfluidic device prepared using standard soft lithography methods.~\cite{xia_soft_1998} The chips were prepared with polydimethylsiloxane (Momentive RTV615 A) mixed with $10$\ \% cross-linking agent (Momentive RTV615 B) . The reticulated elastomers were treated under O$_2$ plasma (Femto Science CUTE) to be covalently bonded to a glass coverslip. The chips consisted of a single microchannel with dimensions $\{L,w,h\} = \{8.8\  \SI{}{\centi\meter}, 180\ \SI{}{\micro\meter}, 18\  \SI{}{\micro\meter}\}$, with the long dimension consisting of several U-turns. Flows were driven by a pressure controller (Fluigent MFCS-4C) allowing for pressure drops $\Delta P \leq1000\ \mathrm{mbar}$. 
\begin{figure*}
\includegraphics[scale=1]{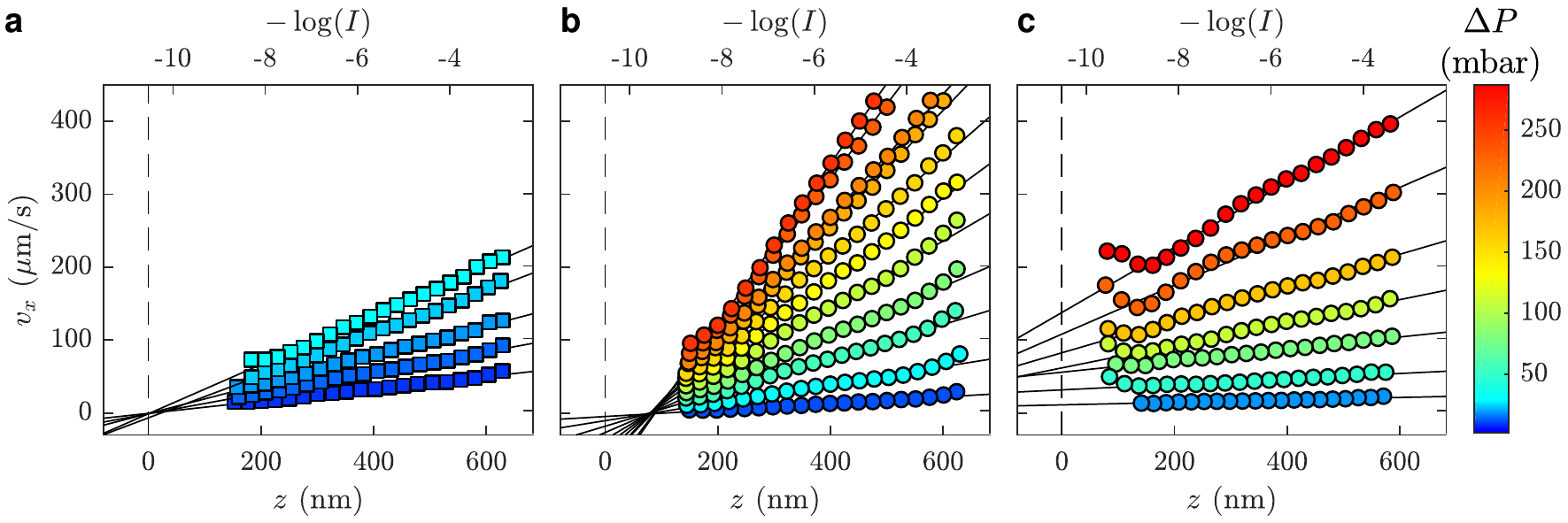} 
\caption{Mean velocity of fluorescent tracers in the flow-direction ($x$) as a function of $-\log(I)$ or $z = \Pi\log(I_0/I)$, for different driving pressures (see the universal color bar at right). The liquids used were (\textbf{a}) water, (\textbf{b}) PAM(2082k) with concentration 2.0\ \SI{}{\milli\gram\per\milli\liter}, and (\textbf{c}) PAM(817k$^{[-]}$) with concentration 0.5\ \SI{}{\milli\gram\per\milli\liter}. Solid straight lines are linear fits of the velocity profiles and dashed lines represent the wall position, taken as the unique intersection point of the water profiles for (\textbf{a}) and as similarly measured in corresponding water flows for (\textbf{b}) and (\textbf{c}).
}
\label{velocity_profiles}
\end{figure*}

For each polymer solution tested, reference measurements were performed in water before the polymer solution was injected into the same microfluidic chip. The TIRF velocimetry measurements were complimented with bulk rheology measurements (Anton Paar, MCR 302) performed in continuous-shear mode in a Couette cell. Strain sweeps were performed with shear rates in the range $1 \leq \dot{\gamma}\leq1000\ \mathrm{s}^{-1}$ with corresponding measurements of the shear stress, $\sigma$. 

\subsection{TIRF microscopy and image processing}
The TIRF microscopy setup along with alignment procedure are described in Ref.~\cite{selvin2008vitro} For our system, a $10\times$-expanded, continuous-wave laser (Coherent Sapphire,  $\lambda =488$\,\SI{}{\nano\meter}) was piloted off the optical axis and focused onto the back focal plane of a high-numerical-aperture, high-magnification objective (Leica, oil immersion HCX PL APO, $NA=$\num{1.46}, $M=100\times$). The beam thus exits the objective at an angle which could be measured using the method described in the appendix. Total internal reflection was achieved when the exiting angle was larger than $\theta_\mathrm{c}=\arcsin(n/n_\mathrm{g})$ where $n_\mathrm{g}=1.518$ is the index of refraction of the glass coverslip and $n$ is that of the solution used. These latter indices were all measured using a refractometer (Atago PAL-RI). Under TIRF conditions particles in the near-surface region emit ~\cite{axelrod_total_1984,kihm_near-wall_2004} a fluorescence intensity 
\begin{align}
I(z)=I_0\exp(-z/\Pi)\ ,
\label{evanescentdecay}
\end{align} 
where $\Pi$, the evanescent decay length, is given by 
\begin{align}
\Pi = \frac{\lambda}{4\pi}\sqrt{\frac{1}{n_\mathrm{g}^{2}\sin^2\theta-n_\mathrm{w}^2}}\ ,
\label{penetrationdepth}
\end{align} 
and $I_0$ is the intensity of a particle at the wall. Thus, TIRFM permits: 1) to observe particles in the first few hundred nanometers of the sample and 2) to infer the distance between the surface and the imaged objects by measuring the fluorescence intensity.

Fluorescence images were collected with an sCMOS-based camera (Andor Neo) under flow in the microfluidic devices described above. For all the tracer-containing flows presented here, \num{2000}-frame movies were recorded at $400$\ \SI{}{\hertz}. Four to seven such videos were recorded for each experimental condition comprising a given polymer solution and driving pressure. This volume of data ensured that at least $10\,000$ particle positions were measured for each experimental realisation of a polymer concentration at a given imposed pressure. Tracers appear as Gaussian-shaped diffraction-limited spots with approximately $10$ pixel (approximately $400$\ \SI{}{\nano\meter}) diameters, giving a precision of \emph{ca.} $10$\ \SI{}{\nano\meter} for the lateral ($x$,$y$) particle positions. The fluorescence intensity ranged from $100$ to $10\,000$ on the 16-bit camera gray scale and were sorted into bins of approximately $0.25$ units of $\log(I)$ (here the natural logarithm), corresponding to a roughly $20$\ nm spread in vertical position for an $80$ nm evanescent decay length. Tracking these binned tracer positions over time allowed for obtention of velocity profiles.  


\section{Results and discussion}
\subsection{Near-wall velocity profiles}
The velocity $v_{x}$, computed as the mean displacement of the tracers within an intensity bin divided by the time elapsed during one frame (2.5 ms), is plotted as a function of $z$ in Fig.~\ref{velocity_profiles}(a), where we use the apparent altitude $z=\Pi\log(I_0/I)$ according to Eq.~\ref{evanescentdecay}. In the appendix, we discuss the limitations~\cite{Sadr2005, Yoda2011, li_near-wall_2015} that may arise from this simple relation. For each pressure used, a linear regression describes the data well. Here, we take the slope of this data as the near-wall shear rate, 
\begin{align}
\dot{\gamma}= \frac{\partial v_\mathrm{x}}{\partial z} = \Pi^{-1}\frac{\partial v_x}{\partial(\log(I_0/I))}\ ,
\label{shearrate}
\end{align} 
which increases as the pressure increases. The linear model applied to the flow profile is justified by the fact the observation window (approximately $600$\ \SI{}{\nano\meter}, \emph{cf.} Fig.~\ref{velocity_profiles}) is small compared to the channel height ($18$\ \SI{}{\micro\meter}). Therefore, a parabolic Poiseuille profile based on Newtonian low-Reynolds-number ($\mathrm{Re}=\rho h v_{x}  / \eta < 10^{-2}$) flow theory is expected. Such a profile can be linearized near the wall with deviations of no more than $3$\ \% for the near-surface region considered here. 

\subsection{Near-wall rheology}
To extract mechanical properties of the fluid from the above-described shear rate measurements, the local shear stress close to the interface is needed. In a rectangular geometry with $h \ll w \ll L$ as satisfied here, a force balance implies that the shear stress is $\sigma (z)=\Delta P(h/2-z)/L$. Therefore, close to the wall one has $\sigma = h\Delta P/2L$. 

Combining the shear rate and the shear stress, we have the viscosity of the fluid which is defined through
\begin{align}
\eta=\frac{\sigma}{\dot{\gamma}} = \frac{h}{2L} \Pi \frac{\Delta P}{\partial v_x/\partial(\log(I_0/I))}\ .
\label{etaexpt}
\end{align}
For the water experiments depicted in Fig.~\ref{velocity_profiles}(a) we obtain an average value for the water viscosity of $\eta_\mathrm{w} = 0.87\pm 0.01$\ \SI{}{\milli \pascal \second}, the error corresponding to the standard deviation of the distribution of the measured viscosities (divided by the square root of the number of measurements) over all of the applied pressures. Systematic errors~\cite{li_near-wall_2015} due to the variations in height of the channels and their length, estimation of the penetration depth and fluctuations in the pressure lead to a uniform shift of this value of $\pm10$\ \% that will be constant for all measurements in the same channel. Considering all of these errors, the quoted value of the water viscosity is consistent with the accepted value\cite{korson_viscosity_1969} at room temperature. Furthermore, we now show that it is possible to use this measurement of the shear-rate-independent water viscosity as a calibration, against which the mechanical properties of other liquids may be compared. 

To this end, we turn our attention to the velocity profiles of neutral PAM(2082k) and anionic PAM(817k$^{[-]}$) that are shown in Fig.~\ref{velocity_profiles}(b) and \ref{velocity_profiles}(c), respectively. As with the profiles for the water flows, the near-wall velocity profiles are well-described by linear regressions, the shear rate obtained through Eq.~\ref{shearrate} also increasing with the pressure drop. We note however, that the pressure drops required to reach a given velocity are larger than those used for water, which is a reflection of the fact that these liquids are more viscous than water. While not shown here, similar trends are seen for water/glycerol mixtures ranging from $0$ to $60$\ wt.\% glycerol.  

\begin{figure*}
\centering
\includegraphics[scale=1]{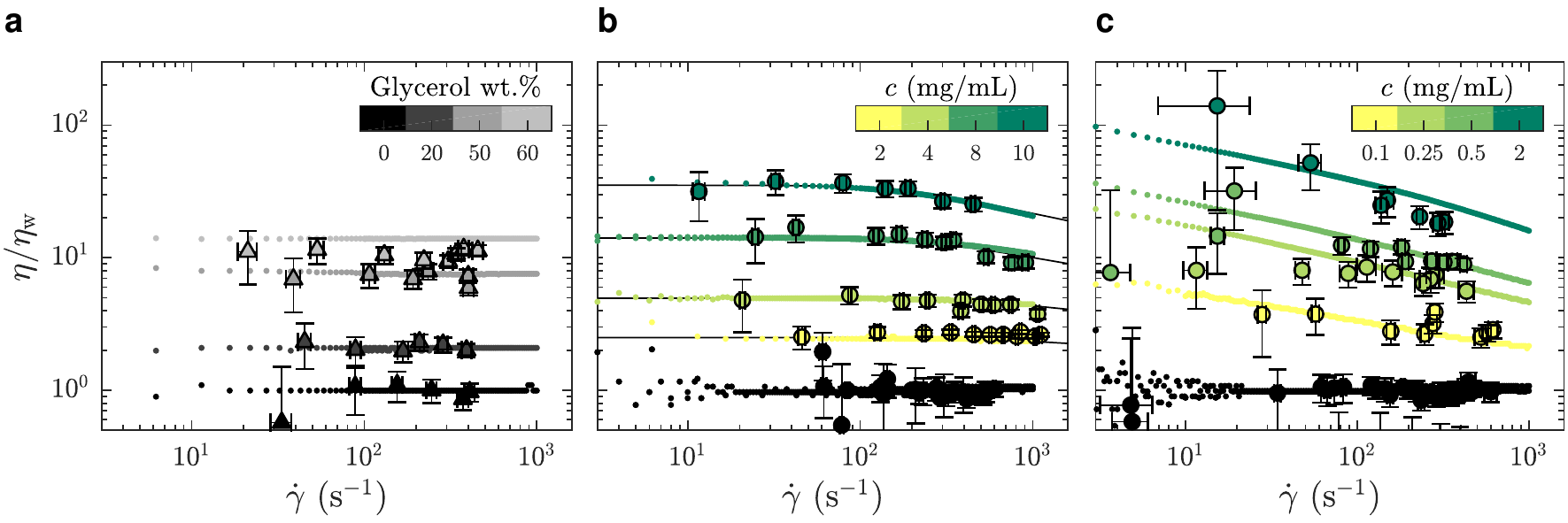} 
\caption{Comparison of the mechanical behaviours of: (\textbf{a}) water/glycerol mixtures;  (\textbf{b}) PAM(2082k) solutions; and (\textbf{c}) PAM(817k$^{[-]}$) solutions. TIRF velocimetry data are plotted in triangles (glycerol mixtures) and circles (semidilute polymer solutions) and rotational rheometry data are plotted using dots; black lines in panel (\textbf{b}) indicate predictions of the universal model of Ref.~\cite{Jouenne2020}. The color bars indicate the concentration of each solution and water is always represented in black.}
\label{rheology}
\end{figure*}

In Fig.~\ref{rheology}(a) is shown the viscosity as a function of the shear rate for such water/glycerol mixtures, obtained through Eq.~\ref{etaexpt}. These viscosities were normalised by the average measured water viscosity for experiments done in the same chip; the TIRFM-derived data are represented with the triangles. Using the viscosity ratio, $\eta/\eta_\mathrm{w}$, the geometric prefactor of Eq.~\ref{etaexpt} cancels. The remaining ratio of the penetration depths is estimated using Eq.~\ref{penetrationdepth}. These liquids are expected to be Newtonian on the range of shear rates accessed by the near-wall velocimetry experiments, and we indeed find that the near-wall nanovelocimetry results do not vary within error over the shear rate range measured. The small points in Fig.~\ref{rheology}(a) were obtained using bulk rheometry, and it is found within error that the data obtained from near-wall TIRFM velocimetry is in agreement with the constant viscosities measured in the bulk. 

In Figs.~\ref{rheology}(b) and (c) are shown $\eta/\eta_\textrm{w}$ -- for neutral PAM(2082k) and anionic PAM(817k$^{[-]}$) -- as a function of the measured shear rates. As with the data in panel (a), the TIRFM-derived data (larger circles) are normalised by the average, constant, water viscosity through Eq.~\ref{etaexpt}. In contrast to the water/glycerol mixtures, however, these solutions contain at most 0.1\,wt\% polymer in water. With such a small amount of polymer, the indices of refraction are almost identical (no more than $0.2$\ \% difference), and thus, according to Eq.~\ref{etaexpt}, this normalisation cancels all of the experimental length scales on the $y$-axis. 

For the neutral polymers in Fig.~\ref{rheology}(b), we observe shear thinning, \emph{i.e.} a viscosity decreasing with shear rate, at the highest concentrations and the highest shear rates. Associated rotational rheometry data, performed with the same solution and normalised in the same way, are superimposed for comparison. The black lines in Fig.~\ref{rheology}(b) show a universal rheological model~\cite{Jouenne2020} applied to the bulk rheology data, with only one fitting parameter. This parameter is the relaxation time of a polymer chain in dilute solution $\lambda_0 = 1.8\pm 0.2$\,\SI{}{\milli\second}, which corresponds well to the expectation for an $R = 66$ nm coil in water, the Zimm time~\cite{rubinstein_polymer_2003} estimated as $\tau_\mathrm{Z} = 6\pi\eta_\textrm{w} R^3/kT\approx 1.3$\ \SI{}{\milli\second}. The other parameter required by the model, that is the intrinsic viscosity, $[\eta]$, was evaluated as  $0.52\pm 0.03$\ \SI{}{\milli\liter\per\milli\gram} by measuring the bulk viscosity as a function of concentration for $c<c^*\approx2$\,\SI{}{\milli\gram\per\milli\liter} (data not shown). The agreement with the model and the bulk rheology are excellent, as is the agreement between these and the TIRFM-derived rheology data. 

Lastly, the anionic PAM(817k$^{[-]}$) of Fig.~\ref{rheology}(c) shows shear thinning in the bulk (small data points) even at the lowest concentrations and shear rates. These solutions being roughly a factor of ten less concentrated than the neutral-PAM solutions, the viscosity nevertheless decreases by almost a decade in the $10$ to $1000$\ \SI{}{\per\second} range. As with the neutral-chain solutions, we note a good agreement between the TIRFM-derived results (larger circles, Eqs.~\ref{shearrate} and~\ref{etaexpt}) and the bulk rheology (smaller points). In the appendix, we show that the bulk results at the lowest shear rate (roughly $5$\ \SI{}{\per\second}) are in agreement with the classical scaling results~\cite{dobrynin_theory_2005} for the zero-shear values of the viscosity as a function of polymer and electrolyte concentration. 

We conclude this subsection generally by noting that for all the liquids we have investigated here, the rheology in the range $150 \lesssim z \lesssim 600$\ \SI{}{\nano\meter} is consistent with that in the bulk. In the following section, we will use this information to unify our description, in particular, of apparent slip boundary conditions for the unentangled semi-dilute solutions of anionic polymers. 

\subsection{Boundary conditions}
We now turn our attention to the boundary conditions of the flows represented in Fig.~\ref{velocity_profiles}, with a particular attention paid to the polymer solutions. For the data in part (a), all of the velocity profiles extrapolate to zero velocity at the same point, to within $9$\ \SI{}{\nano\meter}, which we take as the glass/water interface. This assumption would be justified if the hydrodynamic boundary condition at the glass/water interface were a no-slip condition, as has been confirmed by many other studies,~\cite{cottin-bizonne_nanorheology_2002,lasne_velocity_2008} including notably TIRFM-based investigations.~\cite{li_near-wall_2015} 

Focusing on the upper horizontal scale in Fig.~\ref{velocity_profiles}(a), and under our no-slip assumption, the dashed vertical line corresponds with the fluorescence intensity of a particle at the wall in water, $I_0^\mathrm{water}$. In Fig.~\ref{velocity_profiles}(b), for a solution of PAM(2082k) with $c = 2.0$\ \SI{}{\milli\gram\per\milli\liter}, we draw the same vertical dashed line centred on the same logarithmic intensity, $I_0^\mathrm{water}$, as for the water flow in the same microfluidic chip and for the same incident laser angle. While the velocity profiles for this polymer-laden flow remarkably extrapolate to zero at the same point on the horizontal axis, there is a shift in the extrapolated value of $I_0$ as compared to $I_0^\mathrm{water}$. 

Such a shifted value of $I_0$ for the polymer-laden flow is consistent with the presence of an adsorbed layer of polymer on the glass-water interface that serves as a no-slip boundary plane. Using the exponential relation for $I(z)$ and the penetration depth, $\Pi$, associated to the angle measured in the same experiment (\emph{cf.} Eqs.~\ref{evanescentdecay} and~\ref{penetrationdepth} as well as the appendix), we estimate the adsorbed layer here to be of thickness $D = 77\pm 6$\ \SI{}{\nano\meter}. Such a value is within order unity of the expected chain size in the bulk solution for PAM(2082k). 

\begin{figure}[t!]
\centering
\includegraphics[scale=1]{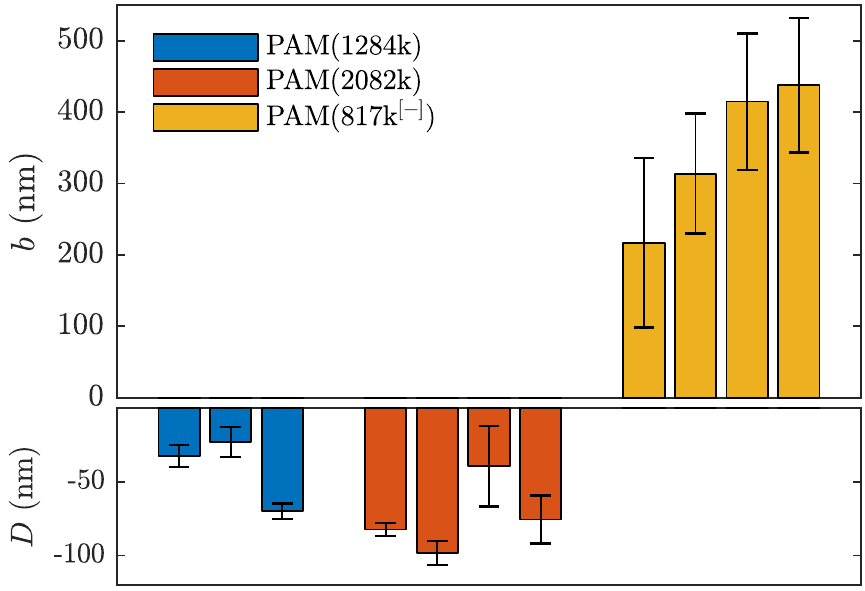} 
\caption{Comparison slip length for different PAM solutions as an average over all imposed pressures, each bar corresponding to a different concentration. From left to right: for PAM(1284k) the concentrations are $c = \{4,12,20\}$\,\SI{}{\milli\gram\per\milli\liter}; for PAM(2082k), $c = \{2,4,8,10\}$\,\SI{}{\milli\gram\per\milli\liter}; and for PAM(817k$^{[-]}$), $c = \{0.1,0.25,0.5,2\}$\,\SI{}{\milli\gram\per\milli\liter}.}
\label{BC}
\end{figure}

In Fig.~\ref{BC}, we show the results of the mean extrapolation lengths for velocity profiles as shown in Fig.~\ref{velocity_profiles} below the glass/liquid interface. The results collect the extrapolation lengths for a range of concentrations for neutral PAMs (two left-most data sets) and for the anionic PAM(817k$^{[-]}$) (right-most data). For the neutral polymers (blue and red), an adsorbed layer of macromolecular size is consistently observed, independent of the concentration of the solution, and growing slightly with molar mass. Specifically the mean adsorbed-layer thickness over the concentrations are $\langle D_{1284}\rangle = 46\pm 11$\ \SI{}{\nano\meter} for PAM(1284k) and $\langle D_{2082} \rangle= 84\pm 9$\ \SI{}{\nano\meter} for PAM(2082k). The ratio of these lengths, $\langle D_{2082}\rangle/\langle D_{1284}\rangle=1.8\pm 0.6$, is, within error, the ratio of the expected chain sizes, $R\sim N^{1/2}$, in semi-dilute solution ~\cite{rubinstein_polymer_2003}, where $N$ is the number of repeat units. Specifically, we have $R_{2082}/R_{1284}\approx(2082/1284)^{1/2} \approx1.3$, the prefactor being a weak power law in the concentration~\cite{rubinstein_polymer_2003} for semi-dilute solutions.

Adsorption of neutral PAM on oxide-based surfaces has been observed for decades in various systems such as nanopores (see \emph{e.g.} Ref.~\cite{bessaies-bey_impact_2018}) or at the surface of colloids.\cite{otsubo_adsorption_1983,lee_adsorption_1989} Additionally, it was predicted that such an adsorbed layer would impact the hydrodynamic boundary condition~\cite{ploehn_self-consistent_1988} in the way depicted in Fig.\ref{introduction_schem}(b) and revealed by the boundary conditions displayed in Figs.~\ref{velocity_profiles} and~\ref{BC}. The parameters influencing the adsorption of chains are notably the nature of the surface and polymer, the molar mass, solvent, pH and temperature; and kinetics is also likely to play a role. While it is beyond the scope of this study to do a detailed analysis of chain adsorption structure and dynamics, the measurements presented here are compatible with the idea that a single chain layer is attached to the wall. These measurements open perspectives on systematic studies of the physicochemical variables described above and possibly to studies of the adsorption dynamics. 

Moving on to a study of the boundary condition for charged chains, we first note that the velocity profiles of PAM(817k$^{[-]}$) in Fig.~\ref{velocity_profiles}(c) extrapolate to zero well below the wall, indicating a slip boundary condition for the flow at the scales measured. As for the neutral PAM, the wall-position measurement was made using a preceding water flow in the same microfluidic chip as for the PAM(817k$^{[-]}$) flow. To complement the measurement of Fig.~\ref{velocity_profiles}(c), we performed identical experiments in independent chips for a range of PAM(817k$^{[-]}$) concentrations. The average extrapolation for each concentration is shown on the right of Fig.~\ref{BC}. Interestingly, the average slip length increases with concentration, while for the adsorbed layer thickness in the case of neutral polymers, there is no clear trend with the concentration. The relatively large error bars for the anionic polymers are due to a definitive trend with the imposed pressure, which we now describe. 

In order to rationalise the results for charged chains in Fig.~\ref{BC}, we apply an apparent slip model that is applicable to systems with a stratified viscosity profile; here we suppose that such stratification is due to an electrostatically-mediated depletion layer. Indeed unlike neutral chains, the interaction between negatively charged PAM and the glass surface is repulsive since glass develops negative charges in aqueous solution.~\cite{behrens_charge_2001} If $d$ is the typical length scale associated with the the low-viscosity depletion layer near the glass/water interface, a simple two-layer model may be proposed. Based on the continuity of stress across the two layers, assuming no slip at the solid/liquid boundary, and considering the geometry shown in Fig.~\ref{introduction_schem}(c) gives the apparent slip length as:
\begin{equation}
b_\mathrm{app} = d\left( \frac{\eta}{\eta_\mathrm{w}}-1\right)\ .
\label{apparent_slip}
\end{equation}
Therefore, rationalising the boundary condition, itself characterising friction at the interface, requires simultaneous knowledge of the rheology. In the context of the velocity profiles measured by TIRFM in Fig.~\ref{velocity_profiles}(c), the term in brackets, usually known as the specific viscosity, of the right hand side of Eq.~\ref{apparent_slip} is evaluated as in Fig.~\ref{rheology}. In this same context, the slip length on the left hand side is evaluated from the intercept of the individual velocity profiles. To test this apparent slip model, a prediction for the depletion layer thickness is thus needed. 

To estimate $d$, we assume that the depletion layer thickness is on the order of the solution's correlation length, $\xi$, defined as the average distance between neighbouring polymer chains in solution, see Fig.~\ref{introduction_schem}(c). This result was predicted by Joanny and co-workers~\cite{joanny_effects_1979} for equilibrium situations, balancing monomeric interactions with the wall and the entropic penalty of the chain which results from its exclusion from the volume near the interface; this typical depletion layer size was confirmed in measurements by Lee and coworkers.~\cite{Lee1991} Following the scaling theory of Dobrynin \emph{et al.}~\cite{Dobrynin1995} for salt-free, semi-dilute polyelectrolyte solutions in a good solvent, we have
\begin{equation}
d \approx \xi \approx a\sqrt{\frac{B}{\tilde{c}a^3}} \sim c^{-1/2} \ .
\label{CorrelationLength}
\end{equation}
Here, $a$ is the monomer size, $\tilde{c}$ is the volumetric monomer concentration, and $B=\left(a \ell_\mathrm{B}^{-1}f^{-2}\right)^{2/7}$ is the ratio of the chain's contour length and its size in dilute solution for an athermal solvent with $f$ the fraction of charged monomers. Finally, the Bjerrum length, $\ell_\mathrm{B} = e^2/\epsilon kT$, is defined as the distance over which the electrostatic energy of two elementary charges ($e$) in a medium of dielectric permittivity, $\epsilon$, is equal to $kT$, the thermal energy at room temperature. For our estimations, we here use $a = 1$\ \SI{}{\nano\meter}, $f=0.1$, and $\epsilon = 80\epsilon_0$ for water with $\epsilon_0$ the vacuum permittivity.  
\begin{figure}[t!]
\centering
\includegraphics[scale=1]{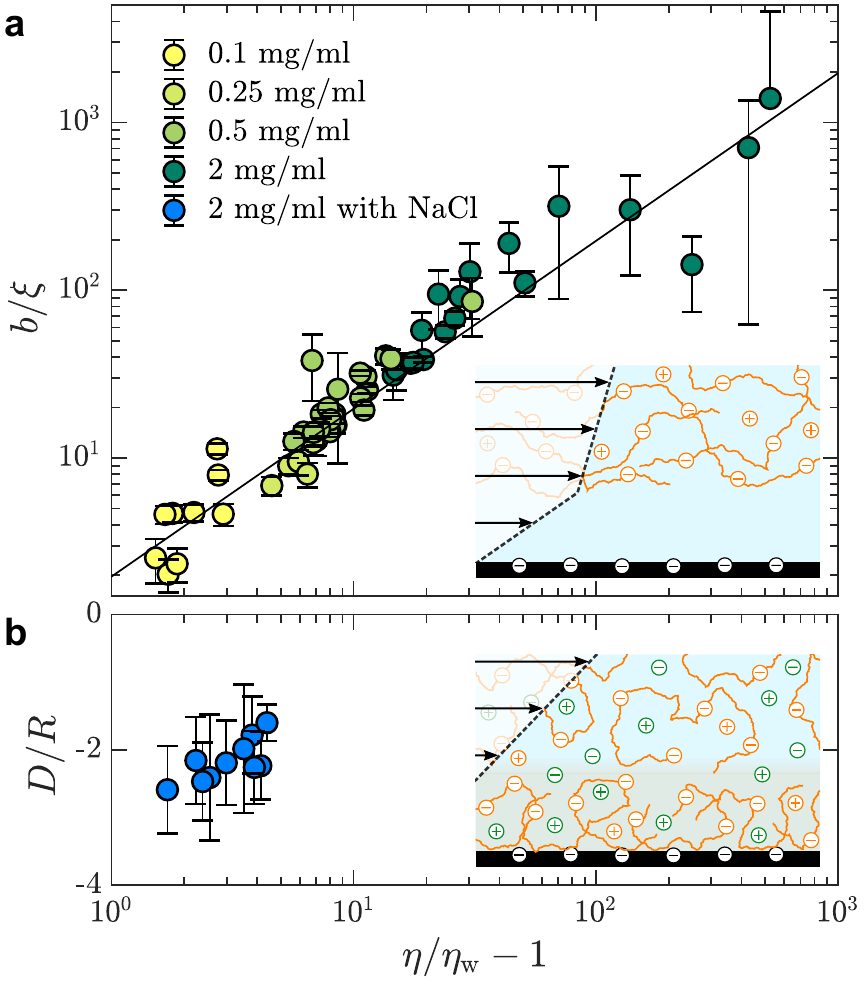} 
\caption{(a) Slip length, normalised by the correlation length, as a function of the specific viscosity for different PAM(817k$^{[-]}$) solutions measured at each imposed pressure. For concentrations $c = \{0.1,0.25,0.5,2\}$\,\SI{}{\milli\gram\per\milli\liter}, we have used $\xi = \{62,39,27,14\}$\,\SI{}{\nano\meter} according to Eq.\ref{CorrelationLength}. The black line shows $b/\xi = A(\eta/\eta_\mathrm{w}-1)$ where $A = 2.0 \pm 0.1$. (\textbf{b}) Adsorbed layer thickness, normalised by the chain size, as a function of specific viscosity for PAM(817k$^{[-]}$) with concentration of NaCl of $c_\mathrm{s} = 28$\,mM. For this solution, we used $R=151$\ \SI{}{\nano\meter} according to ref.~\cite{Dobrynin1995} for chains in a good solvent and with added salt. }
\label{BCneg}
\end{figure}

In Fig.~\ref{BCneg}(a), we test the scaling predictions of Eqs.~\ref{apparent_slip} and~\ref{CorrelationLength} collecting data from all concentrations of PAM(817k$^{[-]}$) and all of the imposed pressures. There we show the normalised slip length as a function of the TIRFM-derived specific viscosity (see Fig.~\ref{rheology}(c)), noting that the specific viscosity changes with the driving pressure. Scaled in this way, we find that all of the data collapse onto a single master curve that is well-described by linear power law in accordance with Eq.~\ref{apparent_slip}. The prefactor is furthermore of order unity, indicating that the depletion layer thickness in these steady shear flows is approximately twice the scaling prediction for the depletion layer thickness at equilibrium, regardless of the shear rate and well into the shear-thinning regime. 

Screening the charges on the anionic polymers and on the glass/water interface is one way to test the validity of our assumption that electrostatic repulsion mediates the apparent slip displayed in Fig.~\ref{BCneg}(a). To this end, we prepared a solution at the highest concentration of anionic polymer (\emph{i.e.} $2.0$\ \SI{}{\milli\gram\per\milli\liter}), in parallel and from the same parent solution as displayed in Fig.~\ref{BCneg}(a), with added salt. The salt concentration, $c_\mathrm{s} = 28$\ mM, was chosen to give roughly ten times the number of charges needed to complement each of the anions on the polymer backbone. In Fig.~\ref{BCneg}(b) are shown the results from this TIRFM-based, combined rheological/boundary condition experiment. As with the unscreened solutions, the viscosity varied as a result of changes in the shear rate attained with changes in the imposed pressure. However, the viscosities are significantly smaller as compared to the unscreened solutions. As for neutral polymers, we find a chain-sized adsorbed layer that varies at most weakly with the imposed pressure (\emph{cf.} Fig.~\ref{velocity_profiles}(b)). 

We thus find that screening the charges on the polymer backbone in a polyelectolyte solution enables to tune the boundary condition in microfluidic contexts with simple control parameters: the polymer concentration and the imposed pressure. By controlling added electrolyte concentration, we are furthermore able to cross the threshold from adsorbed layers of several tens of nanometres, to positive slip length values in the range of one or two hundred nanometres to several microns. 

\section{Conclusion}
Using evanescent wave microscopy and calibrations based on well-known properties of water, we characterise near-surface flows of semi-dilute, neutral and anionic acrylamide-based polymer solutions with macromolecular resolution. The shear rates, extracted from near-wall velocity profiles as a function of the imposed pressure across a microfluidic channel provide the viscosity of the polymer solutions for a wide range of polymer concentration. 
The rheological properties determined within the first micrometer of the surface, are similar those obtained from  bulk rheology measurements, for Newtonian water/glycerol mixtures as well as shear-thinning anionic PAM solutions.

In addition to rheological information, extrapolation of the velocity profiles highlights various boundary conditions for the semi-dilute polymer solutions. The neutral solutions display a chain-sized adsorbed layer wherein the no-slip plane is shifted above the solid/liquid interface; in contrast, anionic polymer solutions exhibit shear-dependant slip lengths. These steady-shear slip lengths are consistent with a bilayer fluid model, wherein viscosity stratification is due to electrostatically-mediated polymer depletion between the glass and polymer. In this model, the near-surface layer has a thickness proportional to the equilibrium value of the correlation length for semi-dilute polyelectrolytes. Lastly, we showed that the apparent slip boundary condition of the anionic PAM solutions is due to the monomer charges: adding salt to screen the charges on the polymer backbone and the glass/water interface, we recover a macromolecular-sized adsorbed layer as for the neutral polymer solutions.

Finally, these experiments show that TIRFM is well-suited to study interfacial phenomena in complex fluids, and especially the underlying mechanism at stake in the friction and slip for semi-dilute polymer solutions. Indeed, it allows to disentangle the two quantities involved in the both phenomena: the local viscosity of the fluid and the friction coefficient for the fluid on the substrate. Our method opens perspectives to describe a rich phenomenology of polymer/substrate interactions under flow.

\section*{Acknowledgements}
The authors thank Mathias Destarac for providing the polymers used here, and warmly thank Stephane Jouenne for insightful discussions. They also benefited from the financial support of CNRS, ESPCI, the Agence Nationale de la Recherche (ANR) under the ENCORE (ANR-15-CE06-005) and CoPinS (ANR-19-CE06-0021) grants, and of the Institut Pierre-Gilles de Gennes (Equipex ANR-10-EQPX-34 and Labex ANR-10-LABX- 31), PSL Research University (Idex ANR-10-IDEX-0001-02). Total is also gratefully acknowledged for financial support.

\section*{Appendix}

\subsection*{Angle measurement}
\begin{figure}[t!]
\includegraphics{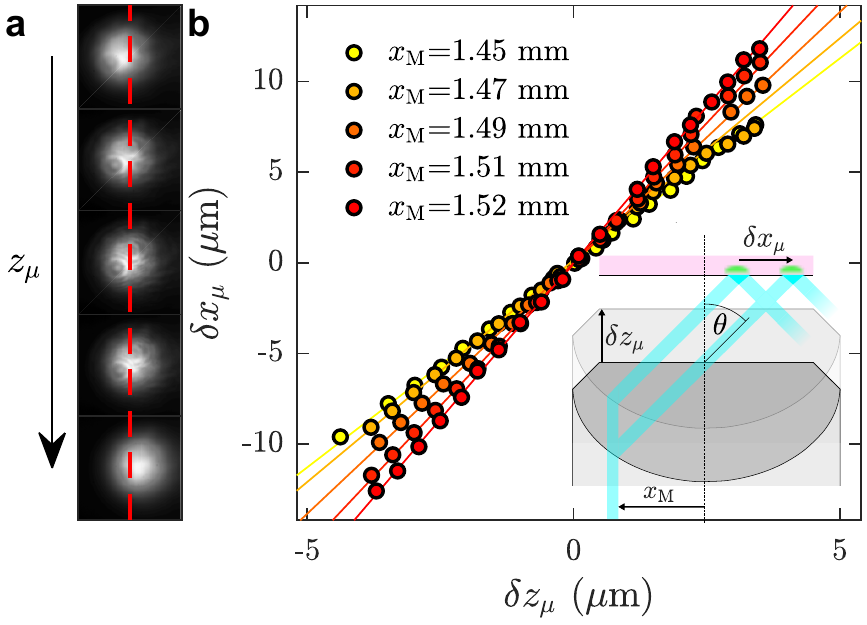} 
\caption{Incident angle calibration. \textbf{(a)} Raw images of the laser beam spot for the \textit{in situ} calibration. Successive images show the spot shifting as imposed $z_\mu$ increases. The red dashed line represents the position of the spot when $\delta z_\mu=$\num{0}. (\textbf{b}) Measured spot position $\delta x_\mu$ as a function of $z_\mu$ for different values of $x_\mathrm{M}$. Straight lines display linear fits of the data.}
\label{angle_1}
\end{figure}

Here we describe the procedure used for measuring the angle of incidence of the laser, which was performed after each data acquisition and in the same microfluidic chip as was used for the investigation of a given fluid of interest. First, a fluorescent dye, typically fluorescein at mass fraction of \num{10}$^{-4}$ in ultrapure water, was injected in the channel. With the laser turned on, a fluorescence spot was observed with full camera resolution (field of view $113 \times 94$\ \SI{}{\micro\meter}, see Fig.~\ref{angle_1}(a)).
Using a piezoelectric objective mount (Physik Instrument, E-662 LVPZT), the objective could be moved vertically in a controlled way, so that the focusing position $z_\mu$ can be tuned. Taking the position for which the fluorescence spot is in focus as a reference, $z_\mu$ is shifted from -5\ \SI{}{\micro\meter} to 5\ \SI{}{\micro\meter} with a 0.4\ \SI{}{\micro\meter} increment, and for each position a snapshot is collected. 
When the focusing plane moved a distance $\delta z_{\mu}$ upward, the fluorescence spot shifted a distance $\delta x_{\mu}$ to the right, as schematically shown in the inset of Fig.~\ref{angle_1}(b); corresponding shifts of the fluorescence spot for different focusing positions are shown in (a). The position $x_{\mu}$ of the spots were determined by 2D-Gaussian-fitting their intensity profiles. In the right of Fig.~\ref{angle_1} are shown $\delta x_{\mu}$ as a function of $\delta z_\mu$ for different values of $x_\mathrm{M}$ corresponding to different values of $\theta$. 

As seen in the figure, for a given value of $x_\mathrm{M}$, $\delta x_\mu$ grows linearly with $\delta z_\mu$, with  a slope $\mathrm{d}\delta x_\mu / \mathrm{d}\delta z_\mu$ increasing with $\theta
$ (see for instance refs.\cite{fish_total_2009,li_near-wall_2015}). The relation $\delta x_\mu = \tan(\theta) \delta z_\mu$ can be derived from the geometry shown in the inset. The experimental results shown in Fig.~\ref{angle_1}(b) indeed display linear relations with slope increasing as $x_\mathrm{M}$ increases, also consistent with the schematic shown in the inset. The angles, $\theta$, predicted from these slopes allow for a prediction of the evanescent penetration depth (\emph{cf.} Eq.~\ref{penetrationdepth}), which are validated by the agreement of our measured water viscosity using TIRFM with accepted values. More directly, and while not shown here, we have verified that the angles thus measured \emph{in situ} are in agreement with the typical method of projecting the laser through a half sphere placed atop the objective~\cite{li_near-wall_2015,zheng_study_2018}, with the horizontal position of the laser measured as a function of $x_\mathrm{M}$. 
\begin{figure}[b!]
\includegraphics{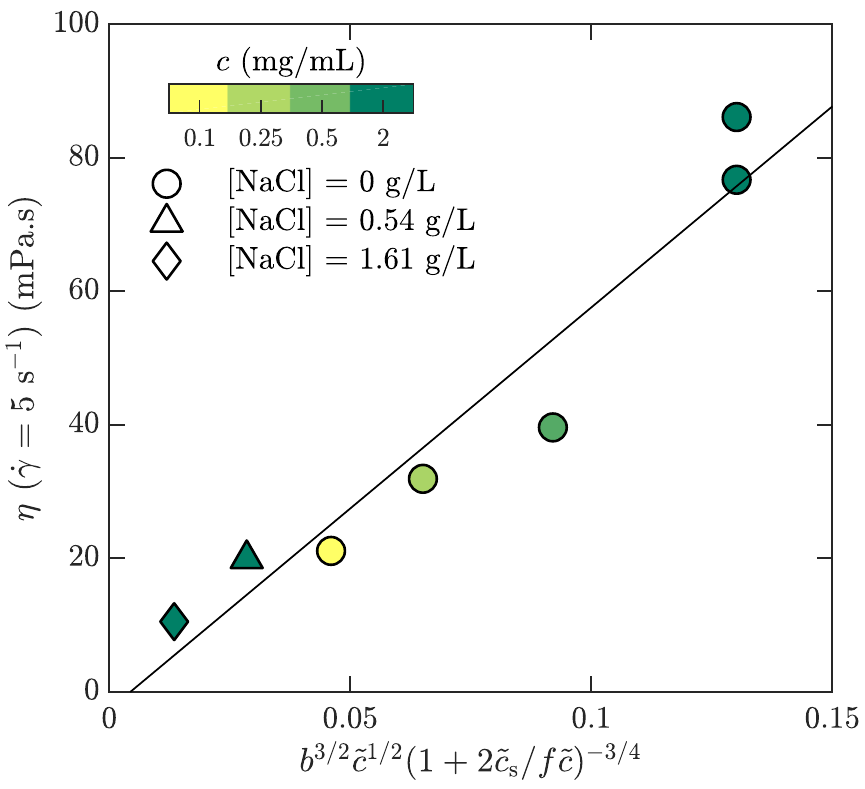} 
\caption{Bulk rheology measurements of the steady-shear viscosity for PAM(817k$^{[-]}$) at a shear rate of 5\ s$^{-1}$, as a function of the scaling variable developed by Dobrynin and coworkers.~\cite{Dobrynin1995}}
\label{anPAMbulk}
\end{figure}
%

\subsection*{Limitations of the particle altitude determination}

The velocimetry data in Fig.~\ref{velocity_profiles} display small deviations from linearity. These are mainly due to optical aberrations: because of the finite depth of field of the objective ($415$ nm), off-focal-plane tracers appear relatively less bright than their in-focus counterparts. Expanding the typical~\cite{zheng_study_2018} Lorentzian multiplicative factor for the intensity as a function of the particle's distance from the focal plane, and considering that the focal plane is typically a few hundred nanometres from the surface, we do not expect that the near-wall linear profile exhibits more than 5\% deviation from linearity due to this defocusing effect. We have furthermore used the signal intensity distribution analyses of ref.~\cite{zheng_study_2018,vilquin_time_2020}, simultaneously fitting the distribution of intensities and velocity profiles and find good agreement with the slopes measured here, within 10\% or so and not varying systematically with the imposed pressure. 

Lastly, we note particular deviations from the linear profile are present at the smallest $z$, especially for part (c) of Fig.~\ref{velocity_profiles}. These deviations arise from Brownian wandering of the tracer particles in the wall-normal direction during the exposure time (2.5 ms),~\cite{Sadr2005, Yoda2011, li_near-wall_2015} coupled with the exponential dependence of the evanescent decay. These effects have been described in detail by several authors, and, in our case, are weighted by a factor proportional to the standard error of the particle displacement within an intensity bin. As a result the data in tail are weighted by a factor $10^{-1}$ (relatively few particles are observed there) as compared to the data in the middle of the range of $z$ and thus do not strongly effect the determination of the shear rate. 

\subsection*{Bulk rheology for PAM(817k$^{[-]}$) -- Comparison to scaling theory}
Here we make a simple comparison between the bulk, steady-shear rheology measured for PAM(817k$^{[-]}$) and the scaling theory of Dobrynin \emph{et al.}~\cite{Dobrynin1995} Without added salt, the specific viscosity, $\eta_\mathrm{r} = \eta/\eta_\mathrm{w}-1$ is expected to scale with the polymer concentration as $\eta_\mathrm{r,0} \sim \tilde{c}^{1/2}$. Including the ionic strength, an additional term counting the (monovalent) ions is needed, such that $\eta_\mathrm{r} = \eta_\mathrm{r,0}\left(1+\frac{2\tilde{c}_\mathrm{s}}{f\tilde{c}}\right)^{-3/4}$, which for large salt concentrations scales as $\eta_\mathrm{r} \sim \tilde{c}^{5/4}$. In Fig.~\ref{anPAMbulk} is shown the relative viscosity measured at the lowest accessed shear rates (5\ \SI{}{\per\second}) as a function variable just described. The data for all of the polymer and salt concentrations fall on a single curve with a straight line describing well the relationship. We thus conclude that the bulk rheology is consistent with standard theories. 


\begin{thebibliography}{0}
\expandafter\ifx\csname natexlab\endcsname\relax\def\natexlab#1{#1}\fi
\expandafter\ifx\csname bibnamefont\endcsname\relax
  \def\bibnamefont#1{#1}\fi
\expandafter\ifx\csname bibfnamefont\endcsname\relax
  \def\bibfnamefont#1{#1}\fi
\expandafter\ifx\csname citenamefont\endcsname\relax
  \def\citenamefont#1{#1}\fi
\expandafter\ifx\csname url\endcsname\relax
  \def\url#1{\texttt{#1}}\fi
\expandafter\ifx\csname urlprefix\endcsname\relax\def\urlprefix{URL }\fi
\providecommand{\bibinfo}[2]{#2}
\providecommand{\eprint}[2][]{\url{#2}}

\end{thebibliography}


\begin{mcitethebibliography}{62}
\providecommand*{\natexlab}[1]{#1}
\providecommand*{\mciteSetBstSublistMode}[1]{}
\providecommand*{\mciteSetBstMaxWidthForm}[2]{}
\providecommand*{\mciteBstWouldAddEndPuncttrue}
  {\def\EndOfBibitem{\unskip.}}
\providecommand*{\mciteBstWouldAddEndPunctfalse}
  {\let\EndOfBibitem\relax}
\providecommand*{\mciteSetBstMidEndSepPunct}[3]{}
\providecommand*{\mciteSetBstSublistLabelBeginEnd}[3]{}
\providecommand*{\EndOfBibitem}{}
\mciteSetBstSublistMode{f}
\mciteSetBstMaxWidthForm{subitem}
{(\emph{\alph{mcitesubitemcount}})}
\mciteSetBstSublistLabelBeginEnd{\mcitemaxwidthsubitemform\space}
{\relax}{\relax}

\bibitem[De~Gennes(1981)]{de_gennes_polymer_1981}
P.~G. De~Gennes, \emph{Macromolecules}, 1981, \textbf{14}, 1637--1644\relax
\mciteBstWouldAddEndPuncttrue
\mciteSetBstMidEndSepPunct{\mcitedefaultmidpunct}
{\mcitedefaultendpunct}{\mcitedefaultseppunct}\relax
\EndOfBibitem
\bibitem[Fu and Santore(1998)]{FU1998}
Z.~Fu and M.~M. Santore, \emph{Colloids and Surfaces A: Physicochemical and
  Engineering Aspects}, 1998, \textbf{135}, 63 -- 75\relax
\mciteBstWouldAddEndPuncttrue
\mciteSetBstMidEndSepPunct{\mcitedefaultmidpunct}
{\mcitedefaultendpunct}{\mcitedefaultseppunct}\relax
\EndOfBibitem
\bibitem[L\'eger \emph{et~al.}(1999)L\'eger, Rapha\"el, and
  Hervet]{leger_surface-anchored_1999}
L.~L\'eger, E.~Rapha\"el and H.~Hervet, in \emph{Polymers in {Confined}
  {Environments}}, ed. S.~Granick, K.~Binder, P.-G. de~Gennes, E.~P. Giannelis,
  G.~S. Grest, H.~Hervet, R.~Krishnamoorti, L.~L\'eger, E.~Manias,
  E.~Rapha{\"e}l and S.-Q. Wang, Springer, Berlin, Heidelberg, 1999, pp.
  185--225\relax
\mciteBstWouldAddEndPuncttrue
\mciteSetBstMidEndSepPunct{\mcitedefaultmidpunct}
{\mcitedefaultendpunct}{\mcitedefaultseppunct}\relax
\EndOfBibitem
\bibitem[Lee \emph{et~al.}(1991)Lee, Guiselin, Lapp, Farnoux, and
  Penfold]{lee1991direct}
L.~Lee, O.~Guiselin, A.~Lapp, B.~Farnoux and J.~Penfold, \emph{Physical review
  letters}, 1991, \textbf{67}, 2838\relax
\mciteBstWouldAddEndPuncttrue
\mciteSetBstMidEndSepPunct{\mcitedefaultmidpunct}
{\mcitedefaultendpunct}{\mcitedefaultseppunct}\relax
\EndOfBibitem
\bibitem[Bessaies-Bey \emph{et~al.}(2018)Bessaies-Bey, Fusier, Harrisson,
  Destarac, Jouenne, Passade-Boupat, Lequeux, d'Espinose~de Lacaillerie, and
  Sanson]{bessaies-bey_impact_2018}
H.~Bessaies-Bey, J.~Fusier, S.~Harrisson, M.~Destarac, S.~Jouenne,
  N.~Passade-Boupat, F.~Lequeux, J.-B. d'Espinose~de Lacaillerie and N.~Sanson,
  \emph{Journal of Colloid and Interface Science}, 2018, \textbf{531},
  693--704\relax
\mciteBstWouldAddEndPuncttrue
\mciteSetBstMidEndSepPunct{\mcitedefaultmidpunct}
{\mcitedefaultendpunct}{\mcitedefaultseppunct}\relax
\EndOfBibitem
\bibitem[Barnes(1995)]{Barnes1995}
H.~A. Barnes, \emph{Journal of Non-Newtonian Fluid Mechanics}, 1995,
  \textbf{56}, 221--251\relax
\mciteBstWouldAddEndPuncttrue
\mciteSetBstMidEndSepPunct{\mcitedefaultmidpunct}
{\mcitedefaultendpunct}{\mcitedefaultseppunct}\relax
\EndOfBibitem
\bibitem[M\"uller‐Mohnssen \emph{et~al.}(1990)M\"uller‐Mohnssen, Weiss, and
  Tippe]{Muller1990}
H.~M\"uller‐Mohnssen, D.~Weiss and A.~Tippe, \emph{Journal of Rheology},
  1990, \textbf{34}, 223--244\relax
\mciteBstWouldAddEndPuncttrue
\mciteSetBstMidEndSepPunct{\mcitedefaultmidpunct}
{\mcitedefaultendpunct}{\mcitedefaultseppunct}\relax
\EndOfBibitem
\bibitem[Navier(1823)]{navier1823memoire}
C.~Navier, \emph{M{\'e}moires de l'Acad{\'e}mie Royale des Sciences de
  l'Institut de France}, 1823, \textbf{6}, 389--440\relax
\mciteBstWouldAddEndPuncttrue
\mciteSetBstMidEndSepPunct{\mcitedefaultmidpunct}
{\mcitedefaultendpunct}{\mcitedefaultseppunct}\relax
\EndOfBibitem
\bibitem[Neto \emph{et~al.}(2005)Neto, Evans, Bonaccurso, Butt, and
  Craig]{neto_boundary_2005}
C.~Neto, D.~R. Evans, E.~Bonaccurso, H.-J. Butt and V.~S.~J. Craig, \emph{Rep.
  Prog. Phys.}, 2005, \textbf{68}, 2859--2897\relax
\mciteBstWouldAddEndPuncttrue
\mciteSetBstMidEndSepPunct{\mcitedefaultmidpunct}
{\mcitedefaultendpunct}{\mcitedefaultseppunct}\relax
\EndOfBibitem
\bibitem[Lauga \emph{et~al.}(2007)Lauga, Brenner, and Stone]{Lauga2007}
E.~Lauga, M.~Brenner and H.~Stone, in \emph{Microfluidics: The No-Slip Boundary
  Condition}, ed. C.~Tropea, A.~L. Yarin and J.~F. Foss, Springer Berlin
  Heidelberg, Berlin, Heidelberg, 2007, pp. 1219--1240\relax
\mciteBstWouldAddEndPuncttrue
\mciteSetBstMidEndSepPunct{\mcitedefaultmidpunct}
{\mcitedefaultendpunct}{\mcitedefaultseppunct}\relax
\EndOfBibitem
\bibitem[Bocquet and Charlaix(2010)]{bocquet_nanofluidics_2010}
L.~Bocquet and E.~Charlaix, \emph{Chem. Soc. Rev.}, 2010, \textbf{39},
  1073--1095\relax
\mciteBstWouldAddEndPuncttrue
\mciteSetBstMidEndSepPunct{\mcitedefaultmidpunct}
{\mcitedefaultendpunct}{\mcitedefaultseppunct}\relax
\EndOfBibitem
\bibitem[Hatzikiriakos(2015)]{hatzikiriakos_slip_2015}
S.~G. Hatzikiriakos, \emph{Soft Matter}, 2015, \textbf{11}, 7851--7856\relax
\mciteBstWouldAddEndPuncttrue
\mciteSetBstMidEndSepPunct{\mcitedefaultmidpunct}
{\mcitedefaultendpunct}{\mcitedefaultseppunct}\relax
\EndOfBibitem
\bibitem[Mhetar and Archer(1998)]{mhetar_slip_1998}
V.~Mhetar and L.~A. Archer, \emph{Macromolecules}, 1998, \textbf{31},
  8607--8616\relax
\mciteBstWouldAddEndPuncttrue
\mciteSetBstMidEndSepPunct{\mcitedefaultmidpunct}
{\mcitedefaultendpunct}{\mcitedefaultseppunct}\relax
\EndOfBibitem
\bibitem[L\'eger \emph{et~al.}(1997)L\'eger, Hervet, Massey, and
  Durliat]{leger_wall_1997}
L.~L\'eger, H.~Hervet, G.~Massey and E.~Durliat, \emph{J. Phys.: Condens.
  Matter}, 1997, \textbf{9}, 7719--7740\relax
\mciteBstWouldAddEndPuncttrue
\mciteSetBstMidEndSepPunct{\mcitedefaultmidpunct}
{\mcitedefaultendpunct}{\mcitedefaultseppunct}\relax
\EndOfBibitem
\bibitem[B\"aumchen \emph{et~al.}(2012)B\"aumchen, Fetzer, Klos, Lessel,
  Marquant, H\"ahl, and Jacobs]{Baumchen2012}
O.~B\"aumchen, R.~Fetzer, M.~Klos, M.~Lessel, L.~Marquant, H.~H\"ahl and
  K.~Jacobs, \emph{J. Phys.: Condens. Matter}, 2012, \textbf{24}, 325102\relax
\mciteBstWouldAddEndPuncttrue
\mciteSetBstMidEndSepPunct{\mcitedefaultmidpunct}
{\mcitedefaultendpunct}{\mcitedefaultseppunct}\relax
\EndOfBibitem
\bibitem[Hatzikiriakos(2012)]{HATZIKIRIAKOS2012}
S.~G. Hatzikiriakos, \emph{Progress in Polymer Science}, 2012, \textbf{37}, 624
  -- 643\relax
\mciteBstWouldAddEndPuncttrue
\mciteSetBstMidEndSepPunct{\mcitedefaultmidpunct}
{\mcitedefaultendpunct}{\mcitedefaultseppunct}\relax
\EndOfBibitem
\bibitem[de~Gennes(1979)]{deGennes1979}
P.-G. de~Gennes, \emph{C.R. Hebd. Seances Acad. Sci.: Ser. B}, 1979,
  \textbf{288}, 219\relax
\mciteBstWouldAddEndPuncttrue
\mciteSetBstMidEndSepPunct{\mcitedefaultmidpunct}
{\mcitedefaultendpunct}{\mcitedefaultseppunct}\relax
\EndOfBibitem
\bibitem[B\"aumchen \emph{et~al.}(2009)B\"aumchen, Fetzer, and
  Jacobs]{baumchen_reduced_2009}
O.~B\"aumchen, R.~Fetzer and K.~Jacobs, \emph{Phys. Rev. Lett.}, 2009,
  \textbf{103}, 247801\relax
\mciteBstWouldAddEndPuncttrue
\mciteSetBstMidEndSepPunct{\mcitedefaultmidpunct}
{\mcitedefaultendpunct}{\mcitedefaultseppunct}\relax
\EndOfBibitem
\bibitem[H\'enot \emph{et~al.}(2018)H\'enot, Drockenmuller, L\'eger, and
  Restagno]{henot_friction_2018}
M.~H\'enot, E.~Drockenmuller, L.~L\'eger and F.~Restagno, \emph{ACS Macro
  Letters}, 2018, \textbf{7}, 112--115\relax
\mciteBstWouldAddEndPuncttrue
\mciteSetBstMidEndSepPunct{\mcitedefaultmidpunct}
{\mcitedefaultendpunct}{\mcitedefaultseppunct}\relax
\EndOfBibitem
\bibitem[H\'enot \emph{et~al.}(2018)H\'enot, Grzelka, Zhang, Mariot, Antoniuk,
  Drockenmuller, L\'eger, and Restagno]{Henot2018}
M.~H\'enot, M.~Grzelka, J.~Zhang, S.~Mariot, I.~Antoniuk, E.~Drockenmuller,
  L.~L\'eger and F.~Restagno, \emph{Phys. Rev. Lett.}, 2018, \textbf{121},
  177802\relax
\mciteBstWouldAddEndPuncttrue
\mciteSetBstMidEndSepPunct{\mcitedefaultmidpunct}
{\mcitedefaultendpunct}{\mcitedefaultseppunct}\relax
\EndOfBibitem
\bibitem[Ilton \emph{et~al.}(2018)Ilton, Salez, Fowler, Rivetti, Aly,
  Benzaquen, McGraw, Rapha\"el, Dalnoki-Veress, and B{\"a}umchen]{Ilton2018}
M.~Ilton, T.~Salez, P.~D. Fowler, M.~Rivetti, M.~Aly, M.~Benzaquen, J.~D.
  McGraw, E.~Rapha\"el, K.~Dalnoki-Veress and O.~B{\"a}umchen, \emph{Nature
  Communications}, 2018, \textbf{9}, 1172\relax
\mciteBstWouldAddEndPuncttrue
\mciteSetBstMidEndSepPunct{\mcitedefaultmidpunct}
{\mcitedefaultendpunct}{\mcitedefaultseppunct}\relax
\EndOfBibitem
\bibitem[Hamieh \emph{et~al.}(2007)Hamieh, Al~Akhrass, Hamieh, Damman,
  Gabriele, Vilmin, Rapha\"el, and Reiter]{HamiehSubstrateDewetting2007}
M.~Hamieh, S.~Al~Akhrass, T.~Hamieh, P.~Damman, S.~Gabriele, T.~Vilmin,
  E.~Rapha\"el and G.~Reiter, \emph{The Journal of Adhesion}, 2007,
  \textbf{83}, 367--381\relax
\mciteBstWouldAddEndPuncttrue
\mciteSetBstMidEndSepPunct{\mcitedefaultmidpunct}
{\mcitedefaultendpunct}{\mcitedefaultseppunct}\relax
\EndOfBibitem
\bibitem[Brochard-Wyart \emph{et~al.}(1996)Brochard-Wyart, Gay, and
  de~Gennes]{Brochard1996}
F.~Brochard-Wyart, C.~Gay and P.-G. de~Gennes, \emph{Macromolecules}, 1996,
  \textbf{29}, 377--382\relax
\mciteBstWouldAddEndPuncttrue
\mciteSetBstMidEndSepPunct{\mcitedefaultmidpunct}
{\mcitedefaultendpunct}{\mcitedefaultseppunct}\relax
\EndOfBibitem
\bibitem[Graham(2011)]{graham_fluid_2011}
M.~D. Graham, \emph{Annu. Rev. Fluid Mech.}, 2011, \textbf{43}, 273--298\relax
\mciteBstWouldAddEndPuncttrue
\mciteSetBstMidEndSepPunct{\mcitedefaultmidpunct}
{\mcitedefaultendpunct}{\mcitedefaultseppunct}\relax
\EndOfBibitem
\bibitem[Chen \emph{et~al.}(2005)Chen, Graham, de~Pablo, Jo, and
  Schwartz]{Chen2005}
Y.-L. Chen, M.~D. Graham, J.~J. de~Pablo, K.~Jo and D.~C. Schwartz,
  \emph{Macromolecules}, 2005, \textbf{38}, 6680--6687\relax
\mciteBstWouldAddEndPuncttrue
\mciteSetBstMidEndSepPunct{\mcitedefaultmidpunct}
{\mcitedefaultendpunct}{\mcitedefaultseppunct}\relax
\EndOfBibitem
\bibitem[Park \emph{et~al.}(2019)Park, Shakya, and King]{Park2019}
S.~J. Park, A.~Shakya and J.~T. King, \emph{Proceedings of the National Academy
  of Sciences}, 2019, \textbf{116}, 16256--16261\relax
\mciteBstWouldAddEndPuncttrue
\mciteSetBstMidEndSepPunct{\mcitedefaultmidpunct}
{\mcitedefaultendpunct}{\mcitedefaultseppunct}\relax
\EndOfBibitem
\bibitem[Cuenca and Bodiguel(2013)]{cuenca_submicron_2013}
A.~Cuenca and H.~Bodiguel, \emph{Phys. Rev. Lett.}, 2013, \textbf{110},
  108304\relax
\mciteBstWouldAddEndPuncttrue
\mciteSetBstMidEndSepPunct{\mcitedefaultmidpunct}
{\mcitedefaultendpunct}{\mcitedefaultseppunct}\relax
\EndOfBibitem
\bibitem[Barraud \emph{et~al.}(2019)Barraud, Cross, Picard, Restagno, L\'eger,
  and Charlaix]{barraud_large_2019}
C.~Barraud, B.~Cross, C.~Picard, F.~Restagno, L.~L\'eger and E.~Charlaix,
  \emph{Soft Matter}, 2019, \textbf{15}, 6308--6317\relax
\mciteBstWouldAddEndPuncttrue
\mciteSetBstMidEndSepPunct{\mcitedefaultmidpunct}
{\mcitedefaultendpunct}{\mcitedefaultseppunct}\relax
\EndOfBibitem
\bibitem[H\'enot \emph{et~al.}(2017)H\'enot, Chennevi\`ere, Drockenmuller,
  L\'eger, and Restagno]{Henot2017}
M.~H\'enot, A.~Chennevi\`ere, E.~Drockenmuller, L.~L\'eger and F.~Restagno,
  \emph{Macromolecules}, 2017, \textbf{50}, 5592--5598\relax
\mciteBstWouldAddEndPuncttrue
\mciteSetBstMidEndSepPunct{\mcitedefaultmidpunct}
{\mcitedefaultendpunct}{\mcitedefaultseppunct}\relax
\EndOfBibitem
\bibitem[Grzelka \emph{et~al.}(2020)Grzelka, Antoniuk, Drockenmuller,
  Chennevi\`ere, L\'eger, and Restagno]{Grzelka2020}
M.~Grzelka, I.~Antoniuk, E.~Drockenmuller, A.~Chennevi\`ere, L.~L\'eger and
  F.~Restagno, \emph{ACS Macro Letters}, 2020, \textbf{9}, 924--928\relax
\mciteBstWouldAddEndPuncttrue
\mciteSetBstMidEndSepPunct{\mcitedefaultmidpunct}
{\mcitedefaultendpunct}{\mcitedefaultseppunct}\relax
\EndOfBibitem
\bibitem[Larson(1999)]{Larson1999}
R.~G. Larson, \emph{The structure and rheology of complex fluids}, Oxford
  University Press, 1999\relax
\mciteBstWouldAddEndPuncttrue
\mciteSetBstMidEndSepPunct{\mcitedefaultmidpunct}
{\mcitedefaultendpunct}{\mcitedefaultseppunct}\relax
\EndOfBibitem
\bibitem[Colby \emph{et~al.}(2007)Colby, Boris, Krause, and
  Dou]{colby_shear_2007}
R.~H. Colby, D.~C. Boris, W.~E. Krause and S.~Dou, \emph{Rheol Acta}, 2007,
  \textbf{46}, 569--575\relax
\mciteBstWouldAddEndPuncttrue
\mciteSetBstMidEndSepPunct{\mcitedefaultmidpunct}
{\mcitedefaultendpunct}{\mcitedefaultseppunct}\relax
\EndOfBibitem
\bibitem[Heo and Larson(2008)]{Heo2008}
Y.~Heo and R.~G. Larson, \emph{Macromolecules}, 2008, \textbf{41},
  8903--8915\relax
\mciteBstWouldAddEndPuncttrue
\mciteSetBstMidEndSepPunct{\mcitedefaultmidpunct}
{\mcitedefaultendpunct}{\mcitedefaultseppunct}\relax
\EndOfBibitem
\bibitem[Jouenne and Levache(2020)]{Jouenne2020}
S.~Jouenne and B.~Levache, \emph{Journal of Rheology}, 2020, \textbf{64},
  1295--1313\relax
\mciteBstWouldAddEndPuncttrue
\mciteSetBstMidEndSepPunct{\mcitedefaultmidpunct}
{\mcitedefaultendpunct}{\mcitedefaultseppunct}\relax
\EndOfBibitem
\bibitem[Kilja\'nski(1989)]{kiljanski_method_1989}
T.~Kilja\'nski, \emph{Rheol Acta}, 1989, \textbf{28}, 61--64\relax
\mciteBstWouldAddEndPuncttrue
\mciteSetBstMidEndSepPunct{\mcitedefaultmidpunct}
{\mcitedefaultendpunct}{\mcitedefaultseppunct}\relax
\EndOfBibitem
\bibitem[Bertula \emph{et~al.}(2019)Bertula, Martikainen, Munne, Hietala,
  Klefstr{\"o}m, Ikkala, and {Nonappa}]{bertula_strain-stiffening_2019}
K.~Bertula, L.~Martikainen, P.~Munne, S.~Hietala, J.~Klefstr{\"o}m, O.~Ikkala
  and {Nonappa}, \emph{ACS Macro Lett.}, 2019, \textbf{8}, 670--675\relax
\mciteBstWouldAddEndPuncttrue
\mciteSetBstMidEndSepPunct{\mcitedefaultmidpunct}
{\mcitedefaultendpunct}{\mcitedefaultseppunct}\relax
\EndOfBibitem
\bibitem[Wang \emph{et~al.}(2011)Wang, Ravindranath, and
  Boukany]{wang_homogeneous_2011}
S.-Q. Wang, S.~Ravindranath and P.~E. Boukany, \emph{Macromolecules}, 2011,
  \textbf{44}, 183--190\relax
\mciteBstWouldAddEndPuncttrue
\mciteSetBstMidEndSepPunct{\mcitedefaultmidpunct}
{\mcitedefaultendpunct}{\mcitedefaultseppunct}\relax
\EndOfBibitem
\bibitem[Nghe \emph{et~al.}(2011)Nghe, Terriac, Schneider, Li, Cloitre,
  Abecassis, and Tabeling]{Nghe2011}
P.~Nghe, E.~Terriac, M.~Schneider, Z.~Z. Li, M.~Cloitre, B.~Abecassis and
  P.~Tabeling, \emph{Lab Chip}, 2011, \textbf{11}, 788--794\relax
\mciteBstWouldAddEndPuncttrue
\mciteSetBstMidEndSepPunct{\mcitedefaultmidpunct}
{\mcitedefaultendpunct}{\mcitedefaultseppunct}\relax
\EndOfBibitem
\bibitem[Read \emph{et~al.}(2014)Read, Guinaudeau, James~Wilson, Cadix,
  Violleau, and Destarac]{Read2014}
E.~Read, A.~Guinaudeau, D.~James~Wilson, A.~Cadix, F.~Violleau and M.~Destarac,
  \emph{Polym. Chem.}, 2014, \textbf{5}, 2202--2207\relax
\mciteBstWouldAddEndPuncttrue
\mciteSetBstMidEndSepPunct{\mcitedefaultmidpunct}
{\mcitedefaultendpunct}{\mcitedefaultseppunct}\relax
\EndOfBibitem
\bibitem[Bessaies-Bey \emph{et~al.}(2019)Bessaies-Bey, Fusier, Hanafi, Zhang,
  Destarac, Jouenne, Passade-Boupat, Lequeux, {d’Espinose de Lacaillerie},
  and Sanson]{BESSAIESBEY2019}
H.~Bessaies-Bey, J.~Fusier, M.~Hanafi, S.~Zhang, M.~Destarac, S.~Jouenne,
  N.~Passade-Boupat, F.~Lequeux, J.-B. {d’Espinose de Lacaillerie} and
  N.~Sanson, \emph{Colloids and Surfaces A: Physicochemical and Engineering
  Aspects}, 2019, \textbf{579}, 123673\relax
\mciteBstWouldAddEndPuncttrue
\mciteSetBstMidEndSepPunct{\mcitedefaultmidpunct}
{\mcitedefaultendpunct}{\mcitedefaultseppunct}\relax
\EndOfBibitem
\bibitem[Dobrynin \emph{et~al.}(1995)Dobrynin, Colby, and
  Rubinstein]{Dobrynin1995}
A.~V. Dobrynin, R.~H. Colby and M.~Rubinstein, \emph{Macromolecules}, 1995,
  \textbf{28}, 1859--1871\relax
\mciteBstWouldAddEndPuncttrue
\mciteSetBstMidEndSepPunct{\mcitedefaultmidpunct}
{\mcitedefaultendpunct}{\mcitedefaultseppunct}\relax
\EndOfBibitem
\bibitem[Xia and Whitesides(1998)]{xia_soft_1998}
Y.~Xia and G.~M. Whitesides, \emph{Angewandte Chemie International Edition},
  1998, \textbf{37}, 550--575\relax
\mciteBstWouldAddEndPuncttrue
\mciteSetBstMidEndSepPunct{\mcitedefaultmidpunct}
{\mcitedefaultendpunct}{\mcitedefaultseppunct}\relax
\EndOfBibitem
\bibitem[Selvin \emph{et~al.}(2008)Selvin, Lougheed, Hoffman, Park, Balci,
  Blehm, and Toprak]{selvin2008vitro}
P.~R. Selvin, T.~Lougheed, M.~T. Hoffman, H.~Park, H.~Balci, B.~H. Blehm and
  E.~Toprak, in \emph{Single-Molecule Techniques: A Laboratory Manual}, Cold
  Spring Harbor Laboratory Press, 2008, ch. 3: In vitro and in vivo FIONA and
  other acronyms for watching molecular motors walk, pp. 37--71\relax
\mciteBstWouldAddEndPuncttrue
\mciteSetBstMidEndSepPunct{\mcitedefaultmidpunct}
{\mcitedefaultendpunct}{\mcitedefaultseppunct}\relax
\EndOfBibitem
\bibitem[Axelrod \emph{et~al.}(1984)Axelrod, Burghardt, and
  Thompson]{axelrod_total_1984}
D.~Axelrod, T.~P. Burghardt and N.~L. Thompson, \emph{Annual Review of
  Biophysics and Bioengineering}, 1984, \textbf{13}, 247--268\relax
\mciteBstWouldAddEndPuncttrue
\mciteSetBstMidEndSepPunct{\mcitedefaultmidpunct}
{\mcitedefaultendpunct}{\mcitedefaultseppunct}\relax
\EndOfBibitem
\bibitem[Kihm \emph{et~al.}(2004)Kihm, Banerjee, Choi, and
  Takagi]{kihm_near-wall_2004}
K.~D. Kihm, A.~Banerjee, C.~K. Choi and T.~Takagi, \emph{Exp Fluids}, 2004,
  \textbf{37}, 811--824\relax
\mciteBstWouldAddEndPuncttrue
\mciteSetBstMidEndSepPunct{\mcitedefaultmidpunct}
{\mcitedefaultendpunct}{\mcitedefaultseppunct}\relax
\EndOfBibitem
\bibitem[Sadr \emph{et~al.}(2005)Sadr, Li, and Yoda]{Sadr2005}
R.~Sadr, H.~Li and M.~Yoda, \emph{Experiments in Fluids}, 2005, \textbf{38},
  90--98\relax
\mciteBstWouldAddEndPuncttrue
\mciteSetBstMidEndSepPunct{\mcitedefaultmidpunct}
{\mcitedefaultendpunct}{\mcitedefaultseppunct}\relax
\EndOfBibitem
\bibitem[Yoda and Kazoe(2011)]{Yoda2011}
M.~Yoda and Y.~Kazoe, \emph{Physics of Fluids}, 2011, \textbf{23}, 111301\relax
\mciteBstWouldAddEndPuncttrue
\mciteSetBstMidEndSepPunct{\mcitedefaultmidpunct}
{\mcitedefaultendpunct}{\mcitedefaultseppunct}\relax
\EndOfBibitem
\bibitem[Li \emph{et~al.}(2015)Li, D'eramo, Lee, Monti, Yonger, Tabeling,
  Chollet, Bresson, and Tran]{li_near-wall_2015}
Z.~Li, L.~D'eramo, C.~Lee, F.~Monti, M.~Yonger, P.~Tabeling, B.~Chollet,
  B.~Bresson and Y.~Tran, \emph{Journal of Fluid Mechanics}, 2015,
  \textbf{766}, 147--171\relax
\mciteBstWouldAddEndPuncttrue
\mciteSetBstMidEndSepPunct{\mcitedefaultmidpunct}
{\mcitedefaultendpunct}{\mcitedefaultseppunct}\relax
\EndOfBibitem
\bibitem[Korson \emph{et~al.}(1969)Korson, Drost-Hansen, and
  Millero]{korson_viscosity_1969}
L.~Korson, W.~Drost-Hansen and F.~J. Millero, \emph{The Journal of Physical
  Chemistry}, 1969, \textbf{78}, 6\relax
\mciteBstWouldAddEndPuncttrue
\mciteSetBstMidEndSepPunct{\mcitedefaultmidpunct}
{\mcitedefaultendpunct}{\mcitedefaultseppunct}\relax
\EndOfBibitem
\bibitem[Rubinstein and Colby(2003)]{rubinstein_polymer_2003}
M.~Rubinstein and R.~Colby, \emph{Polymer {Physics}}, OUP Oxford, 2003\relax
\mciteBstWouldAddEndPuncttrue
\mciteSetBstMidEndSepPunct{\mcitedefaultmidpunct}
{\mcitedefaultendpunct}{\mcitedefaultseppunct}\relax
\EndOfBibitem
\bibitem[Dobrynin and Rubinstein(2005)]{dobrynin_theory_2005}
A.~Dobrynin and M.~Rubinstein, \emph{Progress in Polymer Science}, 2005,
  \textbf{30}, 1049--1118\relax
\mciteBstWouldAddEndPuncttrue
\mciteSetBstMidEndSepPunct{\mcitedefaultmidpunct}
{\mcitedefaultendpunct}{\mcitedefaultseppunct}\relax
\EndOfBibitem
\bibitem[Cottin-Bizonne \emph{et~al.}(2002)Cottin-Bizonne, Jurine, Baudry,
  Crassous, Restagno, and Charlaix]{cottin-bizonne_nanorheology_2002}
C.~Cottin-Bizonne, S.~Jurine, J.~Baudry, J.~Crassous, F.~Restagno and
  E.~Charlaix, \emph{Eur. Phys. J. E}, 2002, \textbf{9}, 47--53\relax
\mciteBstWouldAddEndPuncttrue
\mciteSetBstMidEndSepPunct{\mcitedefaultmidpunct}
{\mcitedefaultendpunct}{\mcitedefaultseppunct}\relax
\EndOfBibitem
\bibitem[Lasne \emph{et~al.}(2008)Lasne, Maali, Amarouchene, Cognet, Lounis,
  and Kellay]{lasne_velocity_2008}
D.~Lasne, A.~Maali, Y.~Amarouchene, L.~Cognet, B.~Lounis and H.~Kellay,
  \emph{Phys. Rev. Lett.}, 2008, \textbf{100}, 214502\relax
\mciteBstWouldAddEndPuncttrue
\mciteSetBstMidEndSepPunct{\mcitedefaultmidpunct}
{\mcitedefaultendpunct}{\mcitedefaultseppunct}\relax
\EndOfBibitem
\bibitem[Otsubo and Umeya(1983)]{otsubo_adsorption_1983}
Y.~Otsubo and K.~Umeya, \emph{Journal of Colloid and Interface Science}, 1983,
  \textbf{95}, 279--282\relax
\mciteBstWouldAddEndPuncttrue
\mciteSetBstMidEndSepPunct{\mcitedefaultmidpunct}
{\mcitedefaultendpunct}{\mcitedefaultseppunct}\relax
\EndOfBibitem
\bibitem[Lee and Somasundaran(1989)]{lee_adsorption_1989}
L.~T. Lee and P.~Somasundaran, \emph{Langmuir}, 1989, \textbf{5},
  854--860\relax
\mciteBstWouldAddEndPuncttrue
\mciteSetBstMidEndSepPunct{\mcitedefaultmidpunct}
{\mcitedefaultendpunct}{\mcitedefaultseppunct}\relax
\EndOfBibitem
\bibitem[Ploehn \emph{et~al.}(1988)Ploehn, Russel, and
  Hall]{ploehn_self-consistent_1988}
H.~J. Ploehn, W.~B. Russel and C.~K. Hall, \emph{Macromolecules}, 1988,
  \textbf{21}, 1075--1085\relax
\mciteBstWouldAddEndPuncttrue
\mciteSetBstMidEndSepPunct{\mcitedefaultmidpunct}
{\mcitedefaultendpunct}{\mcitedefaultseppunct}\relax
\EndOfBibitem
\bibitem[Behrens and Grier(2001)]{behrens_charge_2001}
S.~H. Behrens and D.~G. Grier, \emph{The Journal of Chemical Physics}, 2001,
  \textbf{115}, 6716--6721\relax
\mciteBstWouldAddEndPuncttrue
\mciteSetBstMidEndSepPunct{\mcitedefaultmidpunct}
{\mcitedefaultendpunct}{\mcitedefaultseppunct}\relax
\EndOfBibitem
\bibitem[Joanny \emph{et~al.}(1979)Joanny, Leibler, and
  Gennes]{joanny_effects_1979}
J.~F. Joanny, L.~Leibler and P.~G.~D. Gennes, \emph{Journal of Polymer Science:
  Polymer Physics Edition}, 1979, \textbf{17}, 1073--1084\relax
\mciteBstWouldAddEndPuncttrue
\mciteSetBstMidEndSepPunct{\mcitedefaultmidpunct}
{\mcitedefaultendpunct}{\mcitedefaultseppunct}\relax
\EndOfBibitem
\bibitem[Lee \emph{et~al.}(1991)Lee, Guiselin, Lapp, Farnoux, and
  Penfold]{Lee1991}
L.~T. Lee, O.~Guiselin, A.~Lapp, B.~Farnoux and J.~Penfold, \emph{Phys. Rev.
  Lett.}, 1991, \textbf{67}, 2838--2841\relax
\mciteBstWouldAddEndPuncttrue
\mciteSetBstMidEndSepPunct{\mcitedefaultmidpunct}
{\mcitedefaultendpunct}{\mcitedefaultseppunct}\relax
\EndOfBibitem
\bibitem[Fish(2009)]{fish_total_2009}
K.~N. Fish, \emph{Current Protocols in Cytometry}, 2009, \textbf{50},
  12.18.1--12.18.13\relax
\mciteBstWouldAddEndPuncttrue
\mciteSetBstMidEndSepPunct{\mcitedefaultmidpunct}
{\mcitedefaultendpunct}{\mcitedefaultseppunct}\relax
\EndOfBibitem
\bibitem[Zheng \emph{et~al.}(2018)Zheng, Shi, and Silber-Li]{zheng_study_2018}
X.~Zheng, F.~Shi and Z.~Silber-Li, \emph{Microfluid Nanofluid}, 2018,
  \textbf{22}, 127\relax
\mciteBstWouldAddEndPuncttrue
\mciteSetBstMidEndSepPunct{\mcitedefaultmidpunct}
{\mcitedefaultendpunct}{\mcitedefaultseppunct}\relax
\EndOfBibitem
\bibitem[Vilquin \emph{et~al.}(2020)Vilquin, Bertin, Soulard, Guyard,
  Rapha\"el, Restagno, Salez, and McGraw]{vilquin_time_2020}
A.~Vilquin, V.~Bertin, P.~Soulard, G.~Guyard, E.~Rapha\"el, F.~Restagno,
  T.~Salez and J.~McGraw, \emph{arXiv:2007.08261 [cond-mat, physics:physics]},
  2020\relax
\mciteBstWouldAddEndPuncttrue
\mciteSetBstMidEndSepPunct{\mcitedefaultmidpunct}
{\mcitedefaultendpunct}{\mcitedefaultseppunct}\relax
\EndOfBibitem
\end{mcitethebibliography}

\providecommand*{\mcitethebibliography}{\thebibliography}
\csname @ifundefined\endcsname{endmcitethebibliography}
{\let\endmcitethebibliography\endthebibliography}{}


\end{document}